\documentclass[%
 reprint,
 amsmath,amssymb,
 aps,
pra,
]{revtex4-2}

\usepackage{graphicx}
\usepackage{dcolumn}
\usepackage{bm}
\usepackage{hyperref}
\hypersetup{
    colorlinks=true,
    linkcolor=blue,
    citecolor=blue,
    urlcolor=blue,
    }
\usepackage[mathlines]{lineno}
\usepackage{mathtools}
\usepackage{physics}
\usepackage{xr-hyper}
\usepackage{placeins}
\usepackage{multirow}
\usepackage{url}
\usepackage[dvipsnames]{xcolor}
\usepackage{soul}

\makeatletter
\newcommand*{\addFileDependency}[1]{
\typeout{(#1)}
\@addtofilelist{#1}
\IfFileExists{#1}{}{\typeout{No file #1.}}
}\makeatother

\newcommand*{\myexternaldocument}[1]{%
\externaldocument{#1}%
\addFileDependency{#1.tex}%
\addFileDependency{#1.aux}%
}
\newcommand{\apssamp}[1][]{in the main text.}
\myexternaldocument{SI}

\begin{document}

\preprint{APS/123-QED}

\title{Origin of the Fano interference and its tunability with near-field interactions in a guided mode-resonant metasurface}

\author{Amitrajit Nag}
\email{amitrajitnag@iisc.ac.in}
\author{Jaydeep K. Basu}%
 \email{basu@iisc.ac.in}
\affiliation{Department of Physics, Indian Institute of Science, C.V. Raman Road, Bengaluru, India - 560012.}%

\begin{abstract}
Asymmetric resonances emerging from the Fano interference are a well-known phenomenon in fields like atomic physics and grating optics, and they have recently started to gain interest in artificially engineered dielectric, metallic, or composite metasurfaces and metamaterials. The guided mode-resonant metasurface belongs to this class with grating-waveguide responses and shows asymmetric resonances. Here, we have theoretically studied the origin of the resonance, finding out the root of the Fano interference. We have followed the ab initio theory derived from the Feshbach formalism for the electromagnetic scattering. We have numerically simulated the metasurface to obtain different field parameters required for the ab initio theory; in this regard, we have used the multipole decomposition of the scattering fields for the induced moments. Motivated by our recent experiments, we have used a planewave and polarized dipole sources to excite the metasurface, and studied subsequent effects, like the resonance redshift and the resonance linewidth narrowing for the metasurface. These happened due to the change of the excitation, and we have numerically quantified them. The change of the excitation source bears the novelty of the work. Thus, it helps comprehend the experimental observations qualitatively. This work could help explain and explore possibilities to observe asymmetric resonances in metasurfaces for various excitation conditions, and the key results of resonance shift and linewidth narrowing could be significant for precision applications and fundamental optical studies.
\end{abstract}

\maketitle


\section{\label{sec:intro}
Introduction}
The asymmetric resonance lineshape originating from the interference phenomenon, well-known as the Fano interference \cite{UFano}, is now widely observed in optics and nanophotonics systems \cite{WoodAnomaliesGrating, WoodsAnomaly_Book, WoodsAnomaly_SPP, WoodsAnomaly_4, FanoInOptics1, FanoGMRPC, FanoInMetamaterial, FanoReview, Fano_Nanoscale, Fano_Nanoscale_2, PlasmonicSensing_Fano, Fano_Photonics}. In recent times, such resonances have been observed in metasurfaces and metamaterials, including the plasmonic analogue of electromagnetically induced transparency (EIT) \cite{PlasmonicEg1, PlasmonicEg2, PlasmonicEg3, PlasmonicEg4, PlasmonicEg5, RavindraStrongCoupling, RavindraSinglePhoton, EIT1, EIT2, EIT3}. This sharp, asymmetric resonance is sensitive to the local environment changes where the interference occurs; thus, it finds applications in precision refractive index sensing \cite{Fano_RI_Sensing}, optical modulation, switching \cite{Fano_OpticalSwitch}, etc. The optical resonant systems made of subwavelength nanostructures have a broad range of interests for applications and fundamental studies \cite{FanoInNano1, FanoInNano2, FanoInNano3, FanoInNano4, Nonlinear_Fano, Sharp_Fano_THz}. Fano interference observed in these systems needs a detailed experimental and theoretical investigation. Fano interference stems from the interference between a discrete non-radiative (dark) mode and a radiative (bright) mode continuum. The challenge is identifying these modes and their interference to understand the phenomenon at a fundamental level for each system.\\
Optical resonances in the engineered nanostructures are supported by the induced dipole or multipole moments, like the electric dipole, magnetic dipole, and toroidal dipole \cite{Toroidal1, Toroidal2, Toroidal3, Toroidal4, Toroidal5}. The toroidal dipole moment is generated by the induced current loops along the surface of a torus through its meridians. This was introduced in the nuclear and particle physics \cite{zeldovichAnapole}, and has been studied for sensing and optical modulations in plasmonic or graphene-dielectric metasurface systems \cite{ToroidalSensor, ToroidalModulator1, ToroidalModulator2}, but it remained disregarded in natural media due to weak response. However, toroidal dipole moments in the visible and near-infrared frequencies are found in metasurfaces in recent times \cite{TDdielectricMSR1, TDVIS, TDdielctricMSR2}. C. Zhou et al. experimentally demonstrated that toroidal dipole moments contribute to asymmetric resonances of the near-infrared spectrum of silicon-made nanohole metasurface \cite{MainTDFanoPaper} and quantified that by applying the multipole decomposition at the near-field. This work shows that the high refractive-index dielectric metasurfaces support induced multipole moments in a significant amount. These moments generate the mode continuum accessible to the far-field direction and discrete guided modes, which are non-radiative in that far-field direction, and they interfere. This way, the system becomes favorable for the emergence of Fano resonances.
\begin{figure}[t]
    \centering
    \includegraphics[width=0.45\textwidth]{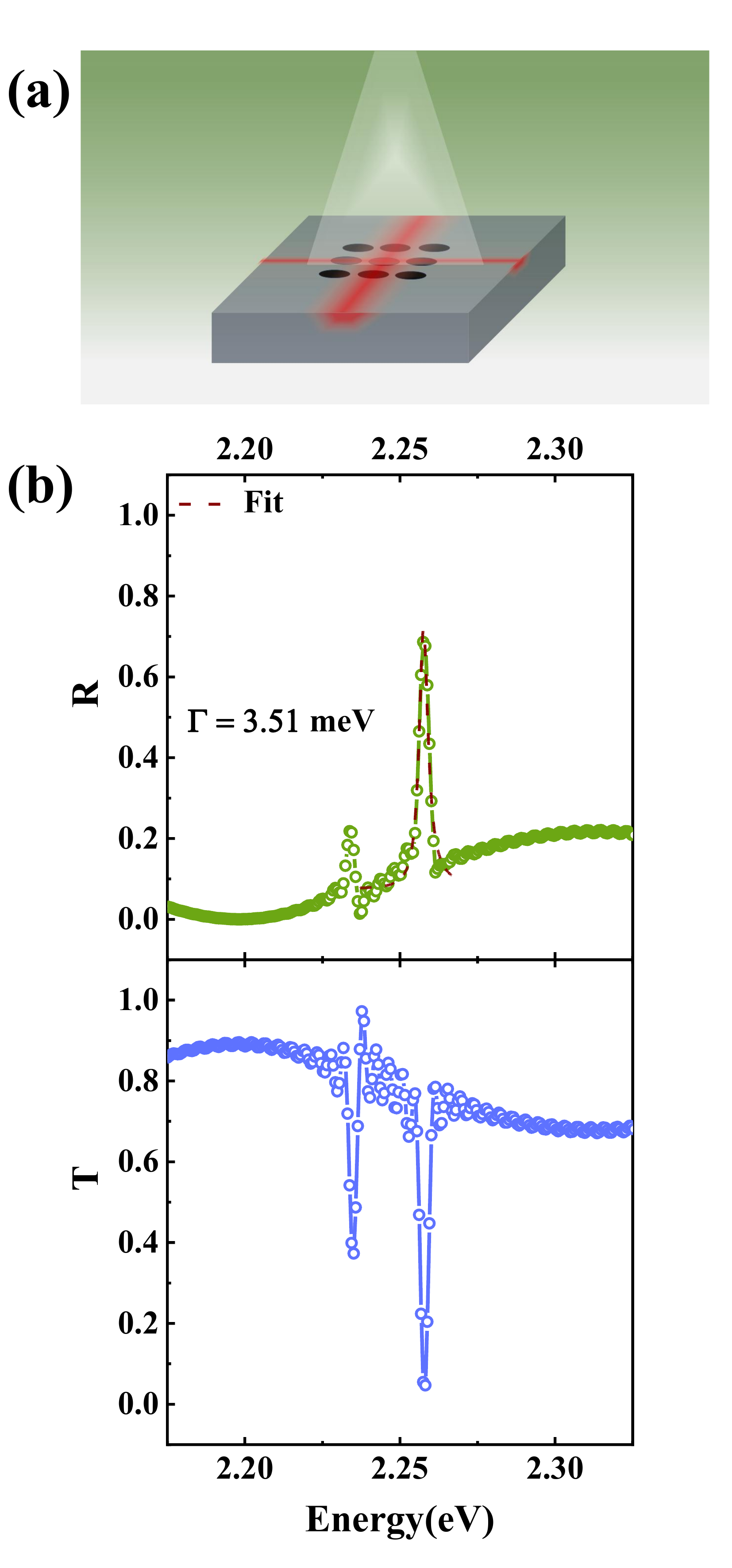}
    \caption{\textbf{(a)} Schematic of nanohole array metasurface excited with an incident plane wave field that generates the asymmetric resonance in the scattering, also gets guided through the underlined waveguide structure as depicted. \textbf{(b)} Simulated far-field spectra of the normalized scattering and transmission of the MSR under planewave excitation.}
    \label{fig:schematic}
\end{figure}
In this work, we aim to understand the near-field interaction of a dielectric guided mode-resonant metasurface (GMR-MSR, or simply MSR) with dipole emitters. In experiments, we have observed an asymmetric resonance when the MSR's resonance was coupled with the quantum emitters' emission mode. The coupled system has shown a resonance redshift and a resonance linewidth narrowing, compared to these corresponding parameters obtained for the planewave excitation of the MSR \cite{OurAdomPaper}. Here, we have theoretically investigated their origin and explained the observations qualitatively. We have followed the ab initio theory of the Fano resonance for electromagnetic scattering from a dielectric
 or metallic object in a dielectric background, based on the Feshbach formalism \cite{FESHBACH, BhatiaTemkin, MalleyGeltman, ProjectionPQ}. Gallinet and Martin have developed this ab initio theory for plasmonic nanostructures \cite{GallinetMartin}. In electrodynamics, a local change of the permittivity can impact the frequency of a resonant cavity. Such resonant open systems manifest the resonance as the complex frequency pole of the corresponding scattering matrix. A small perturbation of the system affects its permittivity and alters the complex pole frequency, resulting in the resonance frequency shift and the change in the linewidth \cite{MSR_perturbation_MaxwellsTheory, MSR_perturbation_PRRes, MSR_perturbation_PRL, ScatMat_MSR_Alu}. The ab initio theory for the electromagnetic scattering explains the perturbation in a specific manner, particularly for metasurface systems showing asymmetric resonance due to the Fano interference. The details of the perturbation theory of the resonant cavity and the derivation of the ab initio theory are given in the supplemental materials (S.M.).
 Next, we will briefly discuss the main equations of the theory used in our work. We have performed numerical simulations following the finite difference time domain (FDTD) method (\textcopyright  FDTD: 3D Electromagnetic Simulator, Lumerical Inc. \cite{Ansys}) to obtain results for the MSR system. We performed simulations on the periodic MSR system and applied two sources, a planewave and a dipole source interacting at the near field of the MSR.

\section{Theoretical methods}
\subsection{\label{sec:theory1}{Brief introduction to the electromagnetic theory of Fano resonances}}
We start the discussion with Maxwell's equations in a dispersing medium where the dielectric scatterer is present as the nanostructure. The dielectric nanohole array GMR-MSR simulated in this work is similar to what has been introduced by D. Rout et al. \cite{DipakGMR} and used in \cite{OurAdomPaper} for numerical calculations. Detailed derivations of the theoretical methods used in this work are presented in the supplemental materials.
The wave equation [Eq.(\ref{eq:S1}) in S.M.] is presented here in terms of the vector wave function of the field, $\ket{\textbf{E}}$. The field is assumed to follow harmonic time dependence $\textbf{E} = \textbf{E}_0e^{-i\omega t}$ throughout the paper.



Feshbach formalism as presented in \cite{GallinetMartin, FESHBACH, BhatiaTemkin, MalleyGeltman, ProjectionPQ} introduces the orthogonal projection operators $P$ and $Q$, which split the field wave function into radiative (bright) and non-radiative (dark) parts, 
and $P\ket{\textbf{E}}$ satisfies the radiation condition in the far field.
Therefore, $Q$ has the orthonormal eigenfunction $\ket{\textbf{E}_d}$, the unique non-radiative mode with eigenvalue $z_d^2$ [Eq.(\ref{eq:S11}) in S.M.]. 
Here, $z_d = \omega_d+i\gamma_d$, $\omega_d$ is the resonance frequency, and $\gamma_d$ is the resonance width due to intrinsic damping. The system should be studied near the resonance frequency $\omega_d$ to enact the operator $Q$. Around the resonance frequency, it can be shown that $Q$ is expanded with respect to its eigenfunctions $\ket{\textbf{E}_d}$ only [Eq.(\ref{eq:S12}) in S.M.] \cite{MalleyGeltman}.
Introducing $P$ and $Q$ in Eq.(\ref{eq:S3}) in S.M. leads to two coupled equations of operators $P$ and $Q$ [Eq.(\ref{eq:S9})-Eq.(\ref{eq:S10}) in S.M.]. Those two coupled equations modify Eq.(\ref{eq:S3}) with a frequency-dependent source term, and yield
\begin{widetext}
    \begin{align}\label{eq:2}
    (\mathcal{M'}_\omega - \omega^2 \textbf{I})P\ket{\textbf{E}} = - P\mathcal{M}_\omega \ket{\textbf{E}_d} \left[\frac{1}{(\omega^2-z_d^2)}\right]\bra{\textbf{E}_d}\mathcal{M}_\omega P\ket{\textbf{E}},
\end{align}
\end{widetext}
$\mathcal{M'}_\omega$, which is modified from the differential operator $\mathcal{M}_\omega$ of the wave equation [Eq.(\ref{eq:S2}) in S.M.], is defined in Eq.(\ref{eq:S17}-\ref{eq:S18}) in S.M.\\
The bright part of the wavefunction $\ket{P\textbf{E}_b}$ satisfies the homogeneous solution of Eq.(\ref{eq:2}),
\begin{equation}\label{eq:3}
    (\mathcal{M'}_\omega - \omega^2 \textbf{I})\ket{P\textbf{E}_b} = 0
\end{equation}

The dyadic Green's function $\mathcal{G}_b$ of Eq.(\ref{eq:3}) is used to solve Eq.(\ref{eq:2}). The solution comes out as,
\begin{gather}\label{eq:4}
    P\ket{\Tilde{\textbf{E}}} = \ket{P\textbf{E}_b}+\frac{\bra{\textbf{E}_d}\mathcal{M}_\omega \ket{P\textbf{E}_b}}{z_d^2 - \omega^2 +\omega_d\Delta}\mathcal{G}_bP\mathcal{M}_\omega Q\ket{\textbf{E}_d},\\
    \text{where,}\;\; \Delta = -\frac{\bra{\textbf{E}_d}\mathcal{M}_\omega P\mathcal{G}_b P\mathcal{M}_\omega \ket{\textbf{E}_d}}{\omega_d}\label{eq:5}
\end{gather}
Here, $\Delta$ is the resonance shift in the frequency for the scattering after the interference occurs between bright and dark mode wavefunctions. Note the negative sign in Eq.(\ref{eq:5}). This signifies the frequency redshift of the scattering resonance. The wavefunction $\ket{\Tilde{\textbf{E}}}$ is related to $\ket{\textbf{E}}$ but does not carry the same asymptotic behavior.\\
Defining the dyadic Green's function $\mathcal{G}_b$ in terms of the wavefunction of the field continuum [Eq.(\ref{eq:S25}) in S.M.], Eq.(\ref{eq:4}) can be simplified to
\begin{equation}\label{eq:6}
    P\ket{\Tilde{\textbf{E}}}=\ket{P\textbf{E}_b}\left[1+\frac{\left|\bra{\textbf{E}_d}\mathcal{M}_\omega \ket{P\textbf{E}_b}\right|^2}{2\omega(z_d^2 - \omega^2 -\omega_d\Delta)} i\right]
\end{equation}
Whereto, the intrinsic damping parameter and resonance width can be defined, brought out of the interference between the field continuum with the discrete part.
\begin{gather}\label{eq:7}
    \Gamma_i = \frac{\left|\bra{\textbf{E}_d}\mathcal{M}_\omega \ket{P\textbf{E}_b}\right|^2 \gamma_d\omega_d}{\omega(z_d^2 - \omega^2 +\omega_d\Delta)^2},\\
    \Gamma = \frac{\left|\bra{\textbf{E}_d}\mathcal{M}_\omega \ket{P\textbf{E}_b}\right|^2}{2\omega(1-\Gamma_i)},\label{eq:8}
\end{gather}
The field term $\ket{\Tilde{\textbf{E}}}$ is introduced to connect $\ket{\textbf{E}}$ with $\ket{\textbf{E}_b}$ in the far field to obtain the same normalization between them (i.e., $\left|P\ket{\textbf{E}}\right|^2=\left|\ket{P\textbf{E}_b}\right|^2$). Therefore, the required relation between $\ket{\textbf{E}}$ and $\ket{\Tilde{\textbf{E}}}$ is,
\begin{equation}\label{eq:9}
    \ket{\textbf{E}}=\frac{\cos{\eta\ket{\Tilde{\textbf{E}}}}}{(1-\Gamma_i)}
\end{equation}
Here $\cot{\eta}=\kappa$, and $\kappa$ is defined as the reduced frequency after the interference,
\begin{gather}\label{eq:10}   
    \kappa = \frac{\omega^2 - \omega_d^2 -\omega_d\Delta}{\Gamma}
\end{gather}
Starting from Eq.(\ref{eq:4}) to Eq.(\ref{eq:10}), all results are the emergence of field overlap between the continuum of the bright field $\ket{P\textbf{E}_b}$ and the discrete dark mode $\ket{\textbf{E}_d}$; the extent of that overlap defines the strength of the interference. That strength tunes the observed effects in the far-field spectrum.\\
Significance of $\eta$ appears in the asymptotic relation between the total field wavefunction $\ket{\textbf{E}}$ and bright field continuum $\ket{\textbf{E}_b}$;
\begin{equation}\label{eq:11}
    P\ket{\textbf{E}}=P\ket{\textbf{E}_b}\exp{i\eta}
\end{equation}
This term rapidly shifts the phase of $\ket{\textbf{E}}$ by $\sim \pi$ in the frequency region $\Gamma$ around the resonance. Since $\eta$ is a complex number, it affects the resonance linewidth of the frequency spectrum in the far field.\\ The resonance redshift, intrinsic damping, resonance width, and the far-field equation, Eq.(\ref{eq:11}), are the key expressions for this work. Our prime results are the comparative study of the numerical results of these expressions obtained for an unpolarized broadband planewave source and a dipole source. The dipole source is embedded in the MSR, sitting at the bottom of the etches of the nanohole array. This scheme was applied in our previous work \cite{OurAdomPaper} for the maximal near-field coupling of the dipole radiation field and the MSR mode. Since the MSR is square-symmetric along its plane, we only need to consider two different polarizations of the dipole source throughout calculations, the vertical and the horizontal polarization with respect to the MSR plane.

\subsection{Multipole decomposition of induced moments in the metasurface to realize the Fano resonance}\label{sec:FanoMoments}
\begin{figure}[t]
    \centering
    \includegraphics[width=0.45\textwidth]{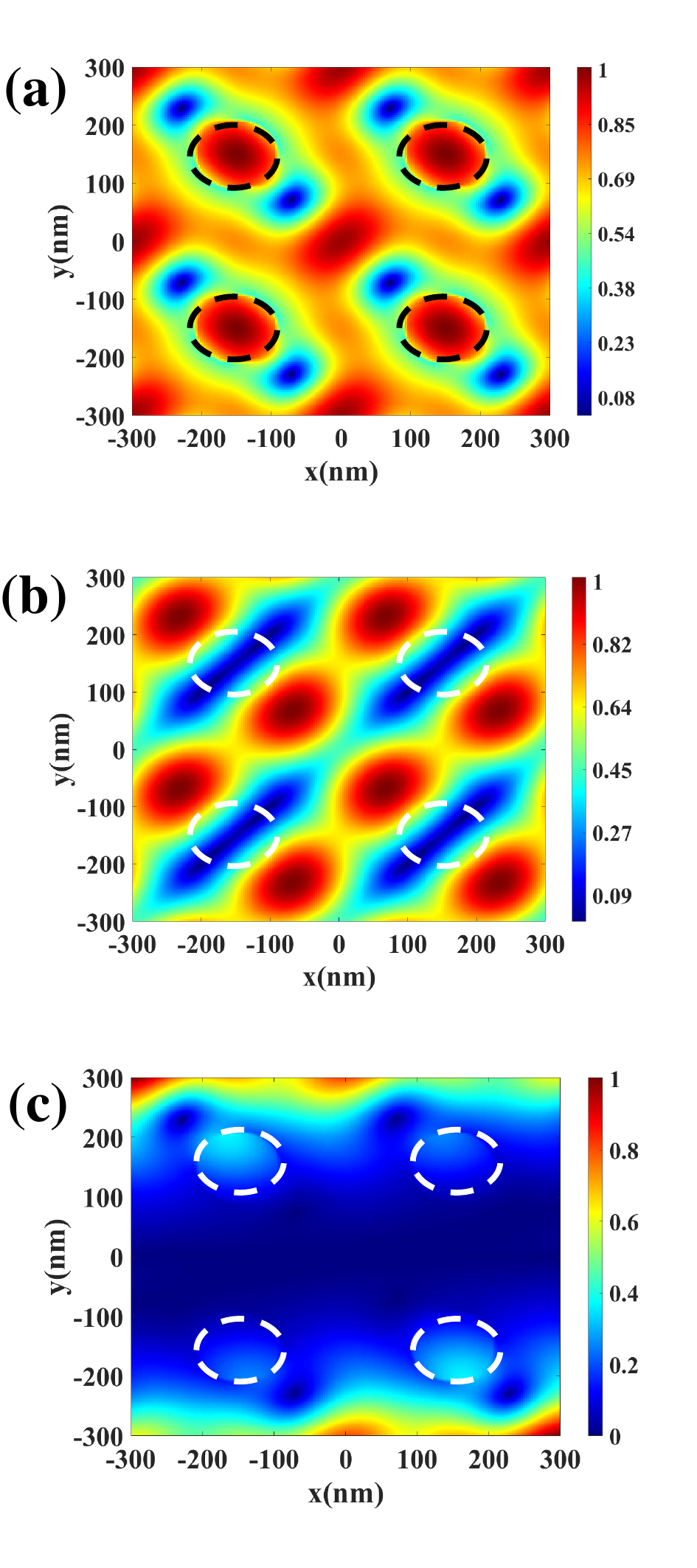}
    \caption{Numerical field profiles of multipoles defined in Eq.(\ref{eq:13}-\ref{eq:15}), obtained from the scattering response inside nanoholes. \textbf{(a)} shows field profile for $\textbf{P}$, \textbf{(b)} shows for $\textbf{M}$, and \textbf{(c)} shows for $\textbf{T}$, respectively, at the resonance frequency. Marked circles describe the positions of MSR etch-nanoholes.}
    \label{fig: fieldprofile}
\end{figure}
Numerical calculations need the decomposition of the MSR scattering response from the induced multipoles in a unit cell.
The multipole moment decomposition of the scattering response for Cartesian coordinates has been defined following these articles \cite{MainTDFanoPaper, Anapole}; the induced current density $\textbf{j}$ in a unit cell of the circular nanohole MSR is given by
\begin{equation}\label{eq:12}
    \textbf{j}=-i\omega\epsilon_0(n^2-1)\textbf{E}
\end{equation}
where $\textbf{E}$ is the local scattering electric field at the MSR unit cell, and $n$ is the refractive index of the MSR material. From this current density $\textbf{j}$, induced multipole moments can be defined;
\begin{gather}\label{eq:13}
    \textbf{P}=\frac{1}{i\omega}\int \textbf{j}d^3r,\\
    \textbf{M}=\frac{1}{2c}\int (\textbf{r}\times\textbf{j})d^3r,\label{eq:14}\\
    \textbf{T}=\frac{1}{10c}\int [(\textbf{r}\cdot\textbf{j})\textbf{r}-2r^2\textbf{j}]d^3r,\label{eq:15}
\end{gather}
$c$ and $\omega$ represent the incident light's speed and angular frequency, respectively. $\textbf{P}$, $\textbf{M}$, and $\textbf{T}$ define the induced electric dipole moment (ED), magnetic dipole moment (MD) and toroidal dipole moment (TD), respectively. Local scattering intensities of these individual moments can also be calculated;
\begin{gather}\label{eq:16}
    I_P=\frac{2\omega^4}{3c^3}\left|\textbf{P}\right|^2,\\
    I_M=\frac{2\omega^4}{3c^3}\left|\textbf{M}\right|^2,\label{eq:17}\\
    I_T=\frac{2\omega^6}{3c^5}\left|\textbf{T}\right|^2 \label{eq:18}
\end{gather}
Following the decomposition [Eq.(\ref{eq:13}-\ref{eq:15})], modes for bright and dark parts of the wavefunction can be separated, as those are used to define the expressions in section \ref{sec:theory1}.

\section{Results}\label{sec:Results}
Figure \ref{fig:schematic}.(a) shows the graphical schematic of the dielectric GMR-MSR. An array of nanoholes is etched over the slab waveguide, which transports the guided mode (shown in red). 
In simulations, we used a square lattice of a circular nanohole array patterned into the silicon nitride (SiN) slab waveguide, supported by an alumina substrate. We used parameters obtained from previous works \cite{OurAdomPaper, DipakGMR}. The nanohole diameter is 120 nm, the periodicity is 300 nm for a unit cell, and the refractive index value of SiN is $n=1.933$.
The corresponding simulated normalized far-field scattering and transmission spectra are shown in Fig.\ref{fig:schematic}.(b).
Figure \ref{fig: fieldprofile} shows the scattering fields decomposed into individual induced contributions as defined by Eq.(\ref{eq:13}-\ref{eq:15}) at the MSR plane within nanoholes. All the field data were recorded with a field monitor placed along the MSR plane inside the nanohole of the unit cell. Further details of the simulations are mentioned in the supplemental materials \cite{supp}.

\begin{figure}[t]
    \centering
    \includegraphics[width=0.45\textwidth]{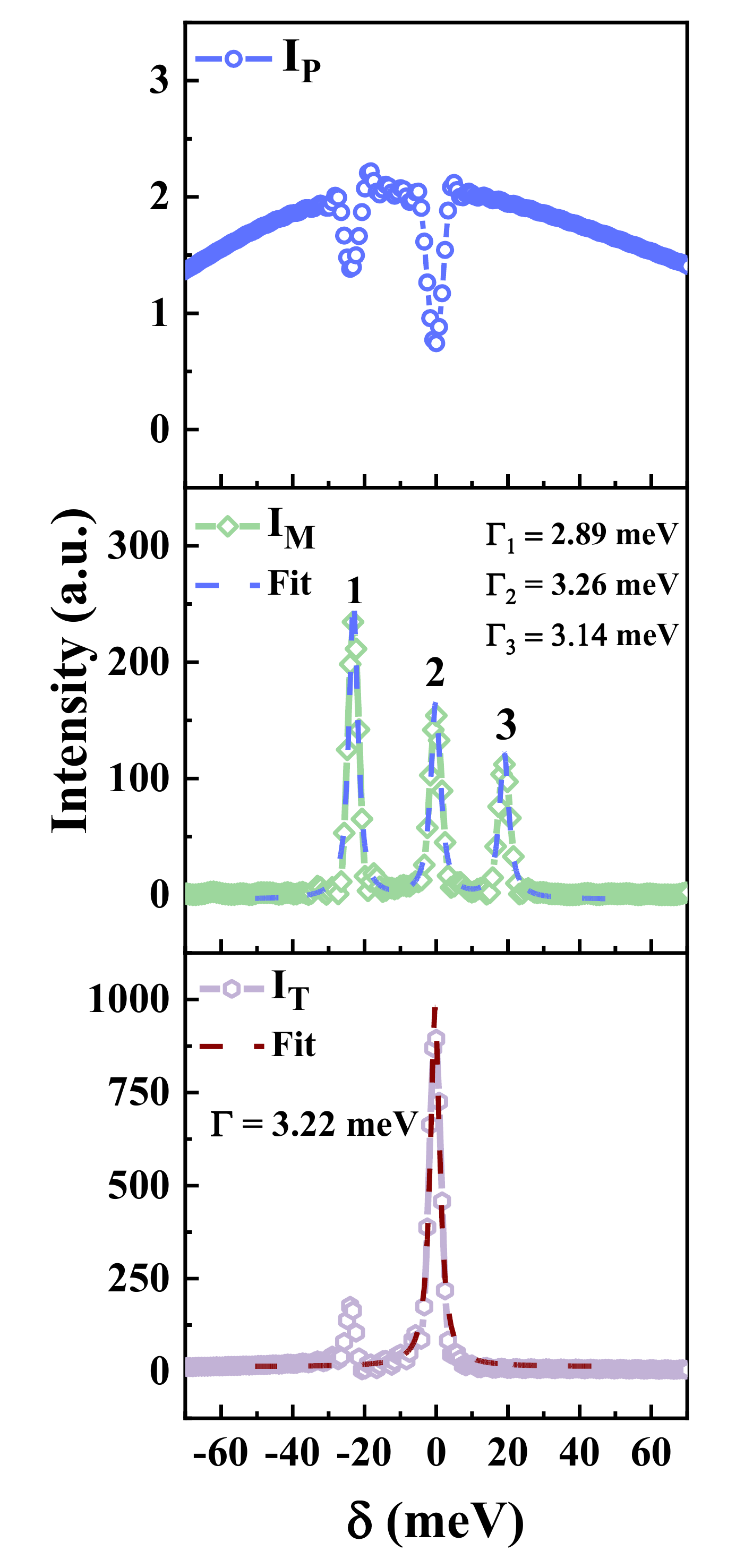}
    \caption{Numerical local scattering intensity plots for respective induced moments of ED, MD, and TD in MSR as expressed in Eq.(\ref{eq:16}-\ref{eq:18}), under planewave excitation.}
    \label{fig:scat_plane}
\end{figure}


First, we used a planewave excitation in the simulation. Figure \ref{fig:scat_plane} shows the numerical local scattering intensity plots for respective induced moments inside the nanohole at the MSR plane, defined by Eq.(\ref{eq:16}-\ref{eq:18}). 
Here, we obtain the field's mode continuum and discrete guided modes in the frequency space. These plots are shown around the main resonance, $\omega_d=2.257$ eV in the reflection and transmission spectra in Fig.\ref{fig:schematic}.(b), and $\delta=\omega-\omega_d$. We have used a dipole source having its transition frequency exactly resonant with this MSR's resonance frequency; therefore, it is convenient to use $\delta$ henceforth. It is evident that the toroidal dipole moment's contribution is dominant among the discrete modes. It can be easily shown that the discrete modes generated by the magnetic and the toroidal moments are the guided modes, and thus, are non-radiative along the normal to the MSR plane, which is the direction of the far-field detection in this work. The details are explained in the supplemental materials \cite{supp}.

In the next simulation, a dipole source was embedded in the MSR nanohole, and that excited the MSR in the near field. The MSR array is square symmetric along its plane; therefore, dipole polarized perpendicular (denoted as vertical or V) and parallel (denoted as horizontal or H) to the MSR plane were used during the simulation. We used  Eq.(\ref{eq:13}-\ref{eq:15}) to yield scattering field responses for the dipole-coupled MSR.
Therefore, numerical results of local scattering responses $\ket{\textbf{PE}_b}$ and $\ket{\textbf{E}_d}$ are now obtained for both the excitations, defined by Eq.(\ref{eq:13}-\ref{eq:15}). Hence, $\Delta$ from Eq.(\ref{eq:5}) can be calculated and compared for the change of the excitation at the resonance frequency $\omega_d$. The data for resonance redshift $\Delta$ is given in Table \ref{table_redshift} for the planewave excitation and the embedded dipole excitations of both polarizations of the MSR.
\begin{table}[!ht]
\caption{\label{table_redshift} Data for the frequency redshift $\Delta$ of the MSR scattering response for MSR excitations with planewave and dipole sources at $\omega_d$.}
\begin{ruledtabular}
\begin{center}
\begin{tabular}{ccc}
\textrm{Excitation}&
\textrm{Interference}&
\multicolumn{1}{c}{\textrm{Resonance redshift}}\\
\textrm{source}&
\textrm{between}&
\multicolumn{1}{c}{\textrm{$\Delta$ (Hz)}}\\
\colrule
Planewave & ED-MD & 29.807\\
 & ED-TD & 11.087\\
V dipole & ED-MD & 8.265\\
 & ED-TD & 700.411\\
H dipole & ED-MD & $3.224\times10^4$\\
 & ED-TD & $2.257\times10^3$\\

\end{tabular}
\end{center}
\end{ruledtabular}
\end{table}

\begin{figure}[t]
    \centering
    \includegraphics[width=0.45\textwidth]{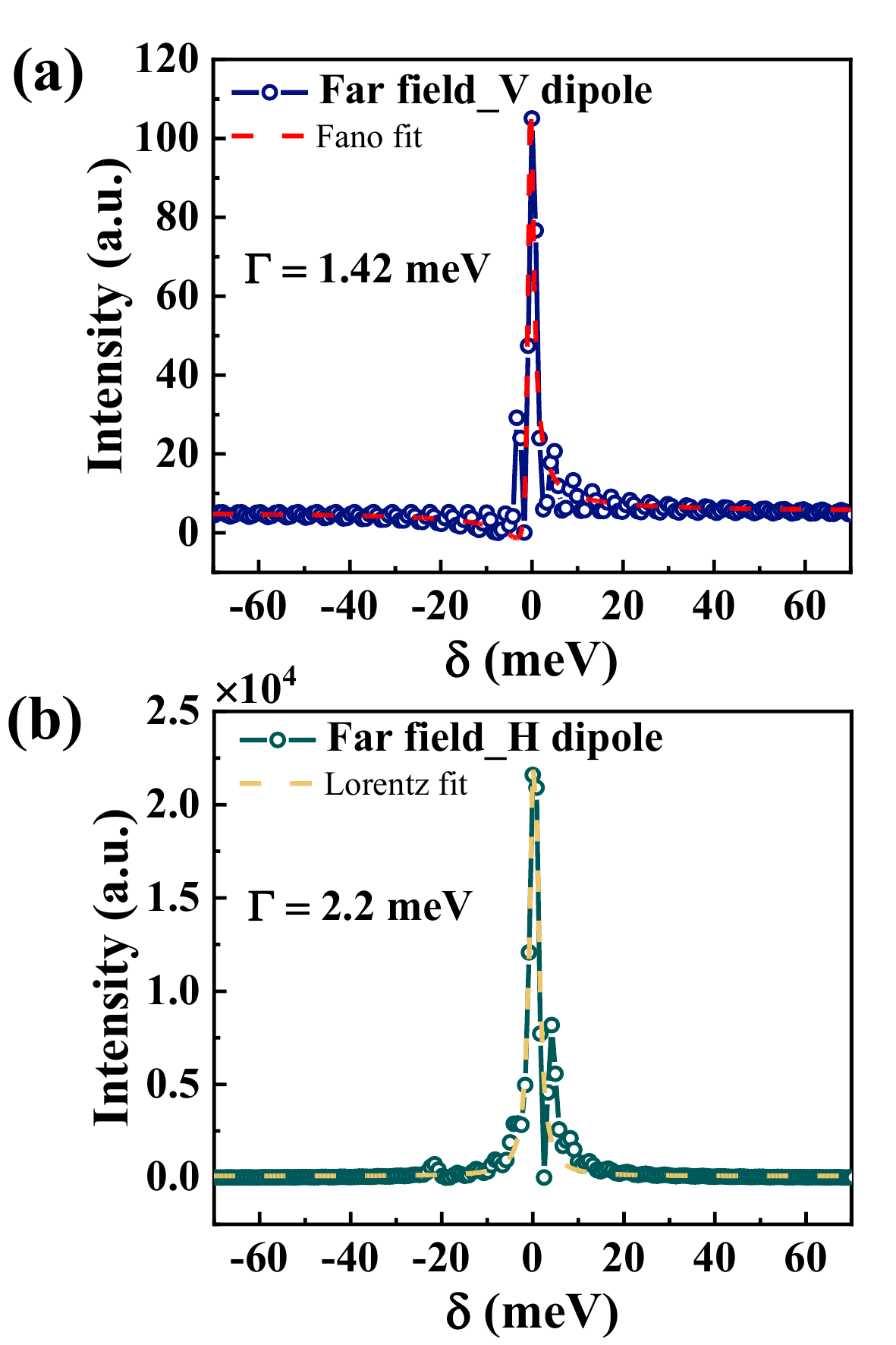}
    \caption{\textbf{(a)}\&\textbf{(b)} Far-field spectrum for the dipole coupled scattering from the MSR for horizontal (H dipole) and vertical (V dipole) polarizations.} 
    \label{fig:farfield}
\end{figure}

An increased magnitude of $\Delta$ obtained for the dipole excitation compared to the planewave would entail the observed resonance redshift. The shift and its amount are defined by Eq.(\ref{eq:4}-\ref{eq:5}); $\omega_d\Delta$ has a unit Hz$^2$. Thus, the shift amount is $\sqrt{\omega_d\Delta}$. Here, for the resonance frequency $\sim 2.257$ eV ($\approx3.429\times10^{15}$ Hz), total $\sqrt{\omega_d\Delta}$ amounts $\approx1\,\mu$eV for the V dipole, and $\approx7.15\,\mu$eV for the H dipole, respectively. These values are obtained by a single dipole interacting with one unit cell of the MSR at periodic boundaries. From Eq.(\ref{eq:5}) and Eq.(\ref{eq:S25}) in S.M., it is evident that the stronger the field overlaps between $\ket{\textbf{PE}_b}$ and $\ket{\textbf{E}_d}$, the stronger the interference would be, eventually increasing the magnitude of the frequency shift.\\
$\Gamma_i$ and $\Gamma$ defined in Eq.(\ref{eq:7}-\ref{eq:8}) also increase due to the increase in the interference. 
However, the far-field response after the interference shows the resonance linewidth narrowing compared to planewave excitation. Figure \ref{fig:farfield}.(a) \& (b) show the far-field scattering response of the MSR defined by Eq.(\ref{eq:11}), for the embedded dipole source of both polarizations. Resonance linewidth reduces $\sim60\%$ for the V dipole and $\sim37\%$ for the H dipole. The phase term in Eq.(\ref{eq:11}) is responsible for the further narrowing, once the excitation is changed from the planewave to the dipole source interacting at the near field. Since $z_d$ is a complex quantity [Eq.(\ref{eq:S11}) in S.M.], hence the complex input remains in Eq.(\ref{eq:5}) and Eq.(\ref{eq:10}-\ref{eq:11}), that yield $\eta$ as a complex number. Therefore, from Eq.(\ref{eq:11}), the imaginary part of $\eta$ will remain in $\left|P\ket{\textbf{E}}\right|^2$ as $\exp(-2\Im{\eta})$; that term squeezes the mode in the frequency range of $\Gamma$ around the resonance. Hence, an increase in $2\Im(\eta)$ would narrow the resonance linewidth in the far field.

\begin{table*}[!ht]
\caption{\label{table_1} Summary of the Fano asymmetry factor ($q$), far-field resonance linewidth $\Gamma_{ff}$, the reduction of $\Gamma_{ff}$, total 2Im$(\eta)$, total $\Gamma$, their difference and the reduction factor. Here, N.A. means that it does not apply to the respective term.}
\begin{ruledtabular}
\begin{center}
\begin{tabular}{ccccccc}
\textrm{Excitation}&
\textrm{$q$}&
\multicolumn{1}{c}{\textrm{$\Gamma_{ff}$}}& Reduction & 2Im($\eta$) & $\Gamma$ & Reduction\\
\textrm{source}& &
\multicolumn{1}{c}{\textrm{(meV)}}& (for $\Gamma_{ff})$& &(a.u,)&(between $\Gamma$ and $\eta$)\\
\colrule
Planewave & 0.905 & 3.51 & N.A. & 81.8 & 41.1 & N.A.\\
V dipole & 2.593 & 1.42 & 0.595 & 1471.4 & 532.54 & 0.653\\
H dipole & N.A. & 2.20 & 0.373 & 69020 & 34578.1 & 0.499\\
\end{tabular}
\end{center}
\end{ruledtabular}
\end{table*}

The result of the trade-off between the increment in the resonance width $\Gamma$ and the increment in the magnitude of $2\Im(\eta)$ is observed in the far-field spectra (Fig.\ref{fig:farfield}). From the definition of Eq.(\ref{eq:9}-\ref{eq:10}), it is evident that $\eta$ and $\Gamma$ are connected, and all these terms stem from the interference. Since the far-field linewidth has a term that is twice that of Im($\eta$), it will always be a greater quantity and would reduce the overall linewidth when the interference has a comparative increment. This has been explained with a detailed break-up of values for relevant parameters in Table \textcolor{cyan}{I} in the supplemental materials \cite{supp}.\\
Table \ref{table_1} shows the comprehensive summary of parameters relevant to the far-field spectra. Here, the asymmetry factor $q$ for the Fano resonance (defined in Eq.(\ref{eq:S33}) in S.M.), resonance linewidth, the reduction of linewidth due to change of the source, 2Im$(\eta)$, and $\Gamma$ for each excitation, and their corresponding overall reductions are discussed.  This table summarizes the prime results of the observed Fano interference in the MSR and its resonance lineshape; the increase in $q$ factor, and an overall reduction in the resonance linewidth for the change of the MSR excitation from the planewave to the embedded dipole source, for both polarizations. The close proximity of reduction factors precisely explains the linewidth reduction mechanism of the far-field spectra. Also, $q$ is not applicable for the H-dipole excited far-field scattering since it is a symmetric, Lorentzian spectrum [Fig.\ref{fig:farfield}.(b)]. In our experiments, we have used two types of quantum emitters, one with anisotropic and the other with isotropic dipole orientations. We observed that the coupled MSR system scattering shows an asymmetric resonance for one type of emitter, and a symmetric resonance for others. Figure \ref{fig:farfield}.(a) and (b) explain their reasons.  
\section{Discussion}
In summary, this work showed the emergence of the Fano resonance in an all-dielectric GMR-MSR made of an array of nanoholes in a high refractive index material, SiN. We took the ab initio theory based on Maxwell's equations and Feshbach's formalism to define the Fano interference between the bright and dark modes with respect to the far field. The interfering fields were resolved following the multipole moment decomposition to find out induced moments and their contributions to the local scattering response of the MSR along the MSR plane within the nanohole. The induced ED response was identified as the bright field, and MD and TD responses were identified as dark modes for obvious reasons: fields generated by these moments don't reach the far field along the outcoupling direction of the MSR, which is along the normal to its plane. Fields from the induced moments MD and TD reach the far-field in the azimuthal directions of the MSR plane, thus constituting the guided modes. (Details are discussed in the supplemental materials \cite{supp}.) Numerical simulations helped quantify scattering field responses for different sources and their polarizations, their corresponding interference term, and construct the far-field spectrum based on the theoretical model that contains important information about the resonant scattering lineshape for the dipole-coupled system. From these interference terms and the far-field spectra, we could explain the experimentally observed phenomena of the resonance redshift, the resonance linewidth reduction, and the resonance lineshape, qualitatively. Therefore, we have successfully applied the ab initio theory for the periodic MSR and used a planewave and a near-field interacting dipole source. We found that the interference terms played a central role in controlling all the observables and their numerical values. These results qualitatively explain the experimental observations; however, they can't compare the values of the parameters exactly. In the experiment, a finite size of MSR with multiple sites interacted with numerous quantum emitters, and here, we simulate a unit cell of a periodic MSR interacting with a dipole source.\\
In the supplemental materials, the theoretical framework is shown to account for the shift of the pole frequency of a resonant cavity system. The real and imaginary parts of the complex frequency account for the resonance position and the linewidth, and both change under a small perturbation of the permittivity of the cavity. This can be very simply understood from Eq.\ref{eq: perturb_poleshift_nonmagnetic} in S.M.  This equation can provide a rough estimate for the experimental scenario, where quantum emitters are plunged into the MSR nanoholes for interaction. From the observations of \cite{OurAdomPaper}, if we assume that a CdSe nanoplatelet has a volume $\sim 300$ nm$^3$, and the volume of an MSR nonohole is $\pi\times200\times60^2=2.262\times10^6$ nm$^3$. The change in the permittivity $\Delta\epsilon\approx(6.5081-1)=5.5081$ around the resonance, then, for $\omega_p=2.2575$ eV, the shift $\Delta\omega_p$ will be
\begin{equation*}
    \Delta\omega_p = -\frac{300}{2.262\times10^6}\times5.5081\times2.2575=1.65\times10^{-3}
\end{equation*}
This provides a resonance shift of 1.65 meV, which is in the order of the experimental observation. However, in this work, we only need a dipole source to model the dipole emitter, and not the material like the CdSe quantum emitter; therefore, here, the permittivity change due to the polarization by the near-field source is not comparable to the order of magnitude obtained in inserting a dielectric material.
\section{Conclusion}
We have theoretically explained the origin of the Fano interference in the guided mode-resonant metasurface and shown that the asymmetric resonance is robust for planewave and near-field interacting dipole sources. One can find the application of this theoretical investigation for any similar metasurface systems. This work will highlight the regimes of applicability of plane wave sources and the limitations of their usage when compared to embedded sources. Precise knowledge of the interference term is a key control parameter of the observables, which can be obtained for relevant systems following this work. The linewidth narrowing effect could be of interest to fundamental studies to investigate subnatural linewidths in the presence of Fano resonance \cite{subnatural1, subnatural2}.\\
\section{Acknowledgement}
Authors acknowledge Prof. Girish S. Agarwal, Dept. of Physics and Astronomy, Texas A \& M University, for useful discussion and suggestions regarding furnishing the work. J.K.B. acknowledges funding from the Anusandhan National Research Foundation (ANRF), India, through grant number CRG/2021/003026 and DST, FIST grant.

\FloatBarrier
\onecolumngrid
\appendix
\section{Pole frequency shift due to perturbing a resonant cavity}\label{app: pole_shift_WignerSmith}
A small perturbation to the permittivity (permeability) of the medium would cause a change in the eigenstate of the system, thus causing a resonance frequency shift \cite{MSR_perturbation_MaxwellsTheory, MSR_perturbation_PRL, MSR_perturbation_PRRes}.  The complex eigenfrequency of the mode is given by $\omega_p=\omega_c-i\frac{\kappa}{2}$. Thus, a shift in $\omega_p$ will cause a shift in the resonance position, given by the real term, and a shift in the linewidth, given by the imaginary term. The frequency shift is quantified by the local permittivity change $\Delta\epsilon$ in a small volume $\Delta V$ containing $\Delta\epsilon$ as shown, for a non-magnetic, resonant cavity, which yields the first-order solution of the pole shift under perturbation \cite{Wigner-Smith_nonHermitian}: 
\begin{equation}\label{eq: perturb_poleshift_nonmagnetic}
    \frac{\Delta\omega_p}{\omega_p}=-\int_{\Delta V} \Delta\epsilon(\textbf{r},\omega_p)\textbf{E}(\textbf{r})\cdot\textbf{E}^*(\textbf{r})dV
\end{equation}
Equation stems from the generalized Wigner-Smith lifetime operator $Q$ that is defined by the corresponding scattering matrix of the arbitrary resonant open cavity system. Here, we briefly introduce that theory.\\
Smith, in his paper \cite{Smith_collision_Qmatrix}, introduced a lifetime matrix $\textbf{Q}$, diagonal elements of which can take care of the lifetimes associated with the corresponding energy eigenstates and their interaction channels. He has developed it in terms of the scattering matrix $\hat{S}$, which is a well-developed tool to describe such an interaction and thereafter the collision completely. It is shown that the elastic collision scattering matrix is defined as $\hat{S}=e^{i\eta}$, and applying that in the Wigner time-delay expression given below, defined by the phase $\eta$ of the particle's wave packet having energy $E$
\begin{equation}\label{eq: Wigner-Smith_time_delay}
    Q=\hbar \frac{d\eta}{dE}=-i\hbar\hat{S}^*\frac{d\hat{S}}{dE}
\end{equation}
It was also shown that the expression of Eq.\ref{eq: Wigner-Smith_time_delay} is appropriate for an inelastic scattering event defined by these operators. Therefore, an arbitrary perturbation can be mapped for a generalized Wigner-Smith lifetime matrix operator, $\textbf{Q}$. This is shown to be applicable in the electromagnetic resonant systems \cite{Wigner_Smith_optics, Wigner_Smith_generalized}. There, Equation \ref{eq: Wigner-Smith_time_delay} takes a form of
\begin{equation}\label{eq: generalized_Wigner-Smith}
    Q = -iS^{-1}\partial_\omega S
\end{equation}
The partial derivative is required since the scattering matrix $S$ is defined by $S=S(\omega,\alpha)$. Here, $\omega$ is the angular frequency and $\alpha$ is a relevant system parameter that accounts for the perturbation. Next, it will be shown how the perturbed system can address the frequency shift of the resonance from the generalized Wigner-Smith operator \cite{Wigner_Smith_generalized}.\\
In this regard, the function $f$ that defines the poles of the scattering matrix by its zeros is given as
\begin{gather}\label{eq: Smatrix_pole}
    f(\omega,\alpha)=\det[S^{-1}(\omega,\alpha)],\\
    f(\omega_p,\alpha)=0, \quad \textrm{when $\omega_p$ is the pole frequency.}
\end{gather}
Here, it is assumed that the poles and zeros are of order one. The system parameter $\alpha$ helps quantify the perturbation so that it changes to $\alpha'=\alpha+\Delta\alpha$. For that, we can expand the function $f$. This can be performed within a closed contour $C$ in the complex plane, which is $\omega\in C$, and includes no poles and zeros of the function, except $\omega_p$. However, due to the perturbation, the pole frequency is also changed, and we get a new function that is expanded.
\begin{equation}\label{eq: pole_expansion}
    f(\omega_p',\alpha')=f(\omega,\alpha)+(\omega_p'-\omega)\partial_\omega f(\omega,\alpha)+\Delta\alpha \partial_\alpha f(\omega,\alpha)+...
\end{equation}
We can safely neglect the higher-order terms since the perturbation and the corresponding shifts are small, holding the validity of the theory \cite{MSR_perturbation_MaxwellsTheory}. Therefore, the perturbed system has the new equation of pole $f(\omega_p',\alpha')=0$. Over $C$, $f$ is nonzero. So, we can divide Eq.\ref{eq: pole_expansion} by $f(\omega,\alpha)$. So, it becomes
\begin{equation}\label{eq: perturbed_pole_expansion}
    0=1+(\omega_p'-\omega)\frac{\partial_\omega f}{f}+\Delta\alpha\frac{ \partial_\alpha f}{f}
\end{equation}
Assuming $\xi\in [\omega,\alpha]$, in general, for these parameters, the logarithmic derivative of $f$ is given by
\begin{equation}
    \frac{\partial_\xi f}{f}=\Tr[S\partial_\xi S^{-1}]=\Tr[-S^{-1}\partial_\xi S]
\end{equation}
Equation \ref{eq: perturbed_pole_expansion} thus has poles for its partial derivative terms at $\omega_p'$. Therefore, to solve them, we need to take the help of the residue theorem. Thus, the solutions are
\begin{equation}
    \begin{split}
        \frac{\Delta\alpha}{2\pi i}\oint_C \frac{\partial_\alpha f}{f} d\omega=&\frac{\Delta\alpha}{2\pi i}\frac{1}{i}\oint_C\Tr[-iS^{-1}\partial_\alpha S]d\omega\\
    =&-i\Delta\alpha \mathop{\mathrm{Res}}_{\omega=\omega_p}\Tr[Q_\alpha]
    \end{split}
\end{equation}
Similarly, we get
\begin{equation}\label{eq: residue_omega}
        (\omega_p'-\omega)\frac{1}{2\pi i}\oint_C \frac{\partial_\omega f}{f} d\omega=(\omega_p'-\omega_p)=\Delta\omega_p
\end{equation}
This is because, by definition, the contour encloses only one pole; thus, the residue number is 1. Therefore, taking the residues of Equation \ref{eq: pole_expansion} yields
\begin{equation}
    \begin{split}
        0=&\Delta\omega_p-i\Delta\alpha \mathop{\mathrm{Res}}_{\omega=\omega_p}\Tr[Q_\alpha]\\
        \implies \Delta\omega_p=&i\Delta\alpha \mathop{\mathrm{Res}}_{\omega=\omega_p}\Tr[Q_\alpha]
    \end{split}
\end{equation}
Equation \ref{eq: residue_omega} can be expanded further,
\begin{equation}
    \begin{split}
        &(\omega_p'-\omega)\frac{1}{2\pi i}\oint_C \frac{\partial_\omega f}{f} d\omega\\
        =& \Delta\omega_p\frac{1}{i}\frac{1}{2\pi i}\oint \Tr[-iS^{-1}\partial_\omega S]d\omega\\
        =& -i\Delta\omega_p \mathop{\mathrm{Res}}_{\omega=\omega_p}\Tr[Q_\omega]=\Delta\omega_p\\
        \implies &\mathop{\mathrm{Res}}_{\omega=\omega_p}\Tr[Q_\omega]^{-1}=-i
    \end{split}
\end{equation}
Therefore, since $i=-1/i$,
\begin{equation}\label{eq: Wigner-Smith_Qmatrix_shift}
    \Delta \omega_p = i\Delta\alpha \mathop{\mathrm{Res}}_{\omega=\omega_p}\Tr[Q_\alpha]= -\Delta\alpha \frac{\mathop{\mathrm{Res}}_{\omega=\omega_p}\Tr[Q_\alpha]}{\mathop{\mathrm{Res}}_{\omega=\omega_p}\Tr[Q_\omega]}
\end{equation}
Therefore, the generalized Wigner-Smith operator for the scattering matrix of a resonant cavity can describe the resonance shift.
\section{Detailed derivation of the results of Electromagnetic theory of Fano resonance}\label{app:Derive_FanoTheory}
We start from the wave equation obtained from Maxwell's equation (Derivation is following the references \cite{GallinetMartin, MalleyGeltman, BhatiaTemkin}); here, the wave equation is framed with operators and vectors in bra-ket notation,
\begin{equation}\label{eq:S1}
    \nabla \times \nabla \times \textbf{E}(\textbf{r},\omega) - \frac{\omega^2}{c^2\epsilon(\textbf{r},\omega)}\textbf{E}(\textbf{r},\omega) = 0
\end{equation}
\begin{equation}\label{eq:S2}
    \mathcal{M}_\omega \textbf{E}(\textbf{r}) = \frac{c^2}{\epsilon} \nabla \times \nabla \times \textbf{E}(\textbf{r})
\end{equation}
\begin{equation}\label{eq:S3}
    (\mathcal{M}_\omega - \omega^2 \textbf{I})\ket{\textbf{E}} = 0
\end{equation}

\begin{equation}\label{eq:S4}
    \ket{\textbf{E}} = P\ket{\textbf{E}} + Q\ket{\textbf{E}}
\end{equation}
Therefore, Eq.(\ref{eq:S3}) becomes,
\begin{equation}\label{eq:S5}
    (\mathcal{M}_\omega - \omega^2 \textbf{I})(P\ket{\textbf{E}} + Q\ket{\textbf{E}}) = 0
\end{equation}
These projection operators, as mentioned in the early works \cite{FESHBACH, MalleyGeltman, ProjectionPQ}, divide the wavefunction into two separate parts; here, these operators are projecting the field wavefunction $\ket{\textbf{E}}$ into radiative (bright) and non-radiative (dark) parts. Also, the bright state is a continuum of modes, whereas the dark one is discrete and corresponds to resonances.\\
Projection operators follow the basic relations,
\begin{align}\label{eq:S6}
    P+Q=I, \quad P^2=P, \quad Q^2=Q;
\end{align}
Next, Eq.(\ref{eq:S5}) can be written as a set of two coupled equations following Eq.(\ref{eq:S6});
\begin{align}\label{eq:S7}
    P(\mathcal{M}_\omega - \omega^2 \textbf{I})(P+Q)\ket{\textbf{E}}= 0 \\
    Q(\mathcal{M}_\omega - \omega^2 \textbf{I})(P+Q)\ket{\textbf{E}}= 0 \label{eq:S8}
\end{align}
Next, Eq.(\ref{eq:S7}) is expanded using the orthogonality condition, $PQ=QP=0$ which readily follows from Eq.(\ref{eq:S6}). Eq.(\ref{eq:S8}) can be expanded in a similar manner.
\begin{gather*}
    (P\mathcal{M}_\omega P - \omega^2 \textbf{I} P^2)\ket{\textbf{E}} + P\mathcal{M}_\omega Q\ket{\textbf{E}} - \omega^2 \textbf{I}PQ\ket{\textbf{E}} = 0 \\
    \implies (P\mathcal{M}_\omega P - \omega^2 \textbf{I})P\ket{\textbf{E}} = -P\mathcal{M}_\omega Q\ket{\textbf{E}}
\end{gather*}
Therefore, the set of the coupled equations become,
\begin{align}\label{eq:S9}
    (P\mathcal{M}_\omega P - \omega^2 \textbf{I})P\ket{\textbf{E}} = -P\mathcal{M}_\omega Q\ket{\textbf{E}} \\
    (Q\mathcal{M}_\omega Q - \omega^2 \textbf{I})Q\ket{\textbf{E}} = -Q\mathcal{M}_\omega P\ket{\textbf{E}} \label{eq:S10}
\end{align}

The discrete non-radiative modes can be expanded as an orthonormal eigenfunction of the corresponding projection operator, i.e., $Q\ket{\textbf{E}_d} =\ket{\textbf{E}_d} $, therefore it satisfies
\begin{equation}\label{eq:S11}
    Q\mathcal{M}_\omega Q\ket{\textbf{E}_d} = z_d^2\ket{\textbf{E}_d}, \quad z_d=\omega_d+i\gamma_d
\end{equation}
Where $\omega_d$ corresponds to mode resonance frequency and $\gamma_d$ goes for intrinsic damping.\\
Choice of $P$ and $Q$ is such that, asymptotically the bright field projection of $\ket{\textbf{E}}$, i.e., $P\ket{\textbf{E}}$ is identical to the total wavefunction $\ket{\textbf{E}}$, i.e., $P\ket{\textbf{E}} \rightarrow \ket{\textbf{E}}$. Owing to this, the non-radiative projection, $Q\ket{\textbf{E}}$ vanishes asymptotically, and remains in the Hilbert space.\\
The completeness of $\ket{\textbf{E}_d}$ being the eigenfunction of $Q$ asserts,
\begin{equation}\label{eq:S12}
    Q = \sum \ket{\textbf{E}_d}\bra{\textbf{E}_d}
\end{equation}
Summation is over the set of discrete eigenfunctions. Equation (\ref{eq:S10}) gives
\begin{equation}\label{eq:S13}
    Q\ket{\textbf{E}}=\left(\frac{1}{\omega^2-Q\mathcal{M}_\omega Q}\right)Q\mathcal{M}_\omega P\ket{\textbf{E}}
\end{equation}
using this expression in Eq.(\ref{eq:S9}) and expanding that gives,
    \begin{gather*}
     P\mathcal{M}_\omega P^2\ket{\textbf{E}}-\omega^2 P\ket{\textbf{E}}+P\mathcal{M}_\omega Q\ket{\textbf{E}}=0\\
     \implies P\mathcal{M}_\omega P^2\ket{\textbf{E}}-\omega^2 P\ket{\textbf{E}}+P\mathcal{M}_\omega \left(\frac{1}{\omega^2-Q\mathcal{M}_\omega Q}\right)Q\mathcal{M}_\omega P\ket{\textbf{E}} = 0\\
     \implies \left[P\mathcal{M}_\omega P+P\mathcal{M}_\omega Q \left(\frac{1}{\omega^2-Q\mathcal{M}_\omega Q}\right)Q\mathcal{M}_\omega P-\omega^2P\textbf{I}\right]\ket{\textbf{E}} = 0
\end{gather*}

Where we have used the identities of Eq.(\ref{eq:S6}). This result summarizes as,
\begin{gather} \label{eq:S14}
    (P\mathcal{M}_\omega P + V -E)\ket{\textbf{E}} =0,\\
    V = P\mathcal{M}_\omega Q \left[\frac{1}{(\omega^2-Q\mathcal{M}_\omega Q)}\right]Q\mathcal{M}_\omega P \label{eq:S15}
\end{gather}
Resonance appears as the zero of the denominator of the effective potential function $V$; since $Q\mathcal{M}_\omega Q\ket{\textbf{E}} = z_d^2\ket{\textbf{E}}$, $V$ can be divided and expressed in two parts; part for the frequency near the resonance frequency $z_d$, and the other part for non-resonant scattering;
    \begin{equation}\label{eq:S16}
    \left(P\mathcal{M}_\omega P+\sum_{\omega'\neq \omega_d}P\mathcal{M}_\omega Q\ket{\textbf{E}'} \left[\frac{1}{(\omega^2-\omega'^2)}\right]\bra{\textbf{E}'}Q\mathcal{M}_\omega P + P\mathcal{M}_\omega Q\ket{\textbf{E}_d} \left[\frac{1}{(\omega^2-z_d^2)}\right]\bra{\textbf{E}_d}Q\mathcal{M}_\omega P-\omega^2P\textbf{I}\right)\ket{\textbf{E}} = 0
\end{equation}

Here the summation part corresponds to the non-resonant scattering of the field, and the second part involves the resonance, which is discrete in frequency space. Hence, this equation becomes
    \begin{gather}\label{eq:S17}
    (\mathcal{M'}_\omega - \omega^2 \textbf{I})P\ket{\textbf{E}} = - P\mathcal{M}_\omega Q\ket{\textbf{E}_d} \left[\frac{1}{(\omega^2-z_d^2)}\right]\bra{\textbf{E}_d}Q\mathcal{M}_\omega P\ket{\textbf{E}},\\
    \text{where,} \;\; \mathcal{M'}_\omega = P\mathcal{M}_\omega P + \sum_{\omega'\neq \omega_d}P\mathcal{M}_\omega Q\ket{\textbf{E}'} \left[\frac{1}{(\omega^2-\omega'^2)}\right]\bra{\textbf{E}'}Q\mathcal{M}_\omega P \label{eq:S18}
\end{gather}
Equation (\ref{eq:S17}) leads to the homogeneous solution of the field wave function satisfying,
\begin{equation}\label{eq:S19}
    (\mathcal{M'}_\omega - \omega^2 \textbf{I})P\ket{\textbf{E}_b} = 0
\end{equation}
Eq.(\ref{eq:S19}) constructs the dyadic Green's function $\mathcal{G}_b$ which provides the formal solution of Eq.(\ref{eq:S17}); 
    \begin{equation}\label{eq:S20}
    P\ket{\Tilde{\textbf{E}}} = P\ket{\textbf{E}_b}+\bra{\textbf{E}_d}Q\mathcal{M}_\omega P\ket{\Tilde{\textbf{E}}}\left[\frac{1}{z^2_d-\omega^2}\right] \mathcal{G}_b P\mathcal{M}_\omega Q\ket{\textbf{E}_d}
\end{equation}

The asymptotic behaviour of $\ket{\textbf{E}}$ and $\ket{\Tilde{\textbf{E}}}$ is different. Their relationship will be discussed later. Now, multiplying $\bra{\textbf{E}_d}Q\mathcal{M}_\omega P$ from the left of Eq.(\ref{eq:S20}), we get,
    \begin{gather*}
    \bra{\textbf{E}_d}Q\mathcal{M}_\omega PP\ket{\Tilde{\textbf{E}}} =  \bra{\textbf{E}_d}Q\mathcal{M}_\omega PP\ket{\textbf{E}_b}+\frac{1}{z^2_d-\omega^2}\bra{\textbf{E}_d}Q\mathcal{M}_\omega P\mathcal{G}_bP\mathcal{M}_\omega Q\ket{\textbf{E}_d}\bra{\textbf{E}_d}Q\mathcal{M}_\omega P\ket{\Tilde{\textbf{E}}},\\
    \implies \bra{\textbf{E}_d}Q\mathcal{M}_\omega P\ket{\Tilde{\textbf{E}}} = \bra{\textbf{E}_d}Q\mathcal{M}_\omega P\ket{\textbf{E}_b}+\frac{1}{z^2_d-\omega^2}\bra{\textbf{E}_d}Q\mathcal{M}_\omega P\mathcal{G}_bP\mathcal{M}_\omega Q\ket{\textbf{E}_d}\bra{\textbf{E}_d}Q\mathcal{M}_\omega P\ket{\Tilde{\textbf{E}}},\\
    \implies \bra{\textbf{E}_d}Q\mathcal{M}_\omega P\ket{\Tilde{\textbf{E}}}\left( 1-\frac{1}{z^2_d-\omega^2}\bra{\textbf{E}_d}Q\mathcal{M}_\omega P\mathcal{G}_b P\mathcal{M}_\omega Q\ket{\textbf{E}_d}\right) =\bra{\textbf{E}_d}Q\mathcal{M}_\omega P\ket{\textbf{E}_b},\\
    \implies \bra{\textbf{E}_d}Q\mathcal{M}_\omega P\ket{\Tilde{\textbf{E}}} =\frac{(z^2_d-\omega^2)}{z^2_d-\omega^2 -\bra{\textbf{E}_d}Q\mathcal{M}_\omega P\mathcal{G}_b P\mathcal{M}_\omega Q\ket{\textbf{E}_d}}\bra{\textbf{E}_d}Q\mathcal{M}_\omega P\ket{\textbf{E}_b}
\end{gather*}

Putting this last expression in Eq.(\ref{eq:S20}) that can be re-written,
\begin{gather}\label{eq:S21}
    P\ket{\Tilde{\textbf{E}}} = P\ket{\textbf{E}_b}+\frac{\bra{\textbf{E}_d}\mathcal{M}_\omega P\ket{\textbf{E}_b}}{z_d^2 - \omega^2 -\omega_d\Delta}\mathcal{G}_bP\mathcal{M}_\omega Q\ket{\textbf{E}_d},\\
    \text{where,}\;\; \Delta = \frac{\bra{\textbf{E}_d}\mathcal{M}_\omega P\mathcal{G}_b P\mathcal{M}_\omega \ket{\textbf{E}_d}}{\omega_d}\label{eq:S22}
\end{gather}
Equation(\ref{eq:S22}) is the expression for the resonance shift of resonance position $\omega_d$ due to the field overlap between the scattering continuum $P\ket{\textbf{E}_b}$ and discrete resonance mode $\ket{\textbf{E}_d}$. The negative sign in the denominator of Eq.(\ref{eq:S21}) shows that, there is a red-shift in resonance frequency due to the field coupling of the discrete mode to the mode continuum.
Hence, the complete expression for the field $\ket{\Tilde{\textbf{E}}}$ can be obtained;
    \begin{gather*}
    \ket{\Tilde{\textbf{E}}} = P\ket{\Tilde{\textbf{E}}}+Q\ket{\Tilde{\textbf{E}}},\\
    \implies \ket{\Tilde{\textbf{E}}} = P\ket{\textbf{E}_b}+\frac{\bra{\textbf{E}_d}\mathcal{M}_\omega P\ket{\textbf{E}_b}}{z_d^2 - \omega^2 -\omega_d\Delta}\mathcal{G}_bP\mathcal{M}_\omega Q\ket{\textbf{E}_d}+Q\ket{\Tilde{\textbf{E}}}
\end{gather*}
Using Eq.(\ref{eq:S13}), last step above can be expanded further,
    \begin{gather}\label{eq:S23}
    \ket{\Tilde{\textbf{E}}} = P\ket{\textbf{E}_b}+\frac{\bra{\textbf{E}_d}\mathcal{M}_\omega P\ket{\textbf{E}_b}}{z_d^2 - \omega^2 -\omega_d\Delta}\mathcal{G}_bP\mathcal{M}_\omega Q\ket{\textbf{E}_d}+\frac{1}{\omega^2-z_d^2}\ket{\textbf{E}_d}\bra{\textbf{E}_d}\mathcal{M}_\omega P\ket{\Tilde{\textbf{E}}}
\end{gather}
The last term $\bra{\textbf{E}_d}\mathcal{M}_\omega P\ket{\Tilde{\textbf{E}}}$ has been simplified already above, and has been used in Eq.(\ref{eq:S20}) to define Eq.(\ref{eq:S21}). Putting that in Eq.(\ref{eq:S23}) yields,
    \begin{gather*}
    \ket{\Tilde{\textbf{E}}} = P\ket{\textbf{E}_b}+\frac{\bra{\textbf{E}_d}\mathcal{M}_\omega P\ket{\textbf{E}_b}}{z_d^2 - \omega^2 -\omega_d\Delta}\mathcal{G}_bP\mathcal{M}_\omega Q\ket{\textbf{E}_d}+\frac{(z_d^2 - \omega^2)}{(\omega^2-z_d^2)(z_d^2 - \omega^2 - \omega_d\Delta)}\ket{\textbf{E}_d}\bra{\textbf{E}_d}\mathcal{M}_\omega P\ket{\textbf{E}_b}, 
\end{gather*}
\begin{equation}
    \implies \ket{\Tilde{\textbf{E}}} = P\ket{\textbf{E}_b}+\frac{\bra{\textbf{E}_d}\mathcal{M}_\omega P\ket{\textbf{E}_b}}{z_d^2 - \omega^2 -\omega_d\Delta}\Bigl(\mathcal{G}_bP\mathcal{M}_\omega \ket{\textbf{E}_d}-\ket{\textbf{E}_d}\Bigr)
\end{equation}

The dyadic Green's function $\mathcal{G}_b$ of Eq.(\ref{eq:S19}) can be expanded w.r.t $P\ket{\textbf{E}_b}$ as orthogonal basis modes \cite{GallinetMartin};
\begin{equation}\label{eq:S25}
    \mathcal{G}_b = \frac{1}{2\pi}\int d\omega'\frac{\ket{P\textbf{E}_b(\omega')}\bra{P\textbf{E}_b(\omega')}}{\omega^{'2}-\omega^2}
\end{equation}
The integral of this form [Eq.(\ref{eq:S25})] is solved using complex contour integration as mentioned in Novotny and Hecht's book \cite{Novotny_Hecht_2012};
    \begin{gather*}
    \int d\omega'\frac{\ket{P\textbf{E}_b(\omega')}\bra{P\textbf{E}_b(\omega')}}{\omega^{'2}-\omega^2} = 2\pi i\; \text{Res}\left(\frac{\ket{P\textbf{E}_b(\omega')}\bra{P\textbf{E}_b(\omega')}}{\omega^{'2}-\omega^2}\right)_{\omega'=\omega}\\
    =\frac{\pi i}{\omega}\ket{P\textbf{E}_b(\omega)}\bra{P\textbf{E}_b(\omega)}
\end{gather*}
Here, only positive frequency is considered. Hence, Eq.(\ref{eq:S21}) becomes,
    \begin{equation*}
    P\ket{\Tilde{\textbf{E}}}=\ket{P\textbf{E}_b}+\frac{i}{2\omega}\frac{\bra{\textbf{E}_d}\mathcal{M}_\omega \ket{P\textbf{E}_b}}{z_d^2 - \omega^2 -\omega_d\Delta}\ket{P\textbf{E}_b}\bra{P\textbf{E}_b}P\mathcal{M}_\omega\ket{\textbf{E}_d}
\end{equation*}
\begin{equation}\label{eq:S26}
    P\ket{\Tilde{\textbf{E}}}=\ket{P\textbf{E}_b}\left[1+\frac{\left|\bra{\textbf{E}_d}\mathcal{M}_\omega \ket{P\textbf{E}_b}\right|^2}{2\omega(z_d^2 - \omega^2 +\omega_d\Delta)} i\right]
\end{equation}
Hereon, the intrinsic damping parameter $\Gamma_i$ and the resonance width $\Gamma$ are defined;
\begin{gather}\label{eq:S27}
    \Gamma_i = \frac{\left|\bra{\textbf{E}_d}\mathcal{M}_\omega \ket{P\textbf{E}_b}\right|^2 \gamma_d\omega_d}{\omega(z_d^2 - \omega^2 -\omega_d\Delta)^2},\\
    \Gamma = \frac{\left|\bra{\textbf{E}_d}\mathcal{M}_\omega \ket{P\textbf{E}_b}\right|^2}{2\omega(1-\Gamma_i)},\label{eq:S28}
\end{gather}
and the reduced frequency term $\kappa$;
\begin{gather}\label{eq:S29}   
    \kappa = \frac{\omega^2 - \omega_d^2 +\omega_d\Delta}{\Gamma}
\end{gather}
The actual field function $\ket{\textbf{E}}$ should have the same normalization at far field as $\ket{\textbf{E}_b}$, hence the relation between $\ket{\textbf{E}}$ and $\ket{\Tilde{\textbf{E}}}$ is,
\begin{equation}\label{eq:S30}
    \ket{\textbf{E}}=\frac{\cos{\eta\ket{\Tilde{\textbf{E}}}}}{(1-\Gamma_i)}
\end{equation}
where, 
\begin{equation}\label{eq:S31}
    \cot{\eta}=\kappa
\end{equation}
with the assumptions $\gamma_d\ll\omega_d$, and only first order contributions are considered.\\
Therefore, asymptotically the bright field wave function follows the relation,
\begin{equation}\label{eq:S32}
    P\ket{\textbf{E}}=P\ket{\textbf{E}_b}\exp{i\eta}
\end{equation}
Here $\eta$ is a complex number. The phase of the field wavefunction shifts rapidly by $\sim \pi$ in a frequency range $\Gamma$ around the resonance frequency.\\
Plots for scattering and far-field spectra having asymmetric resonance are fitted with the following expression defined for Fano resonant scattering;
\begin{gather}\label{eq:S33}
    f = f_0 + h\frac{(\epsilon+q)^2}{(1+\epsilon^2)}, \quad \epsilon=\frac{2(E-E_0)}{\Gamma_R}
\end{gather}
where $h$ parameterizes the amplitude, $q$ the asymmetry factor, $E_0$ is the resonance peak position, and $\Gamma_R$ is the usual resonance width obtained with this fit.
\section{Brief description of the Simulation setup}
We designed a SiN (refractive index was set at 1.933 \cite{DipakGMR}) slab waveguide of 400 nm thickness, and the designed MSR is a square lattice array of nanoholes etched in the waveguide. The slab waveguide is supported by an alumina substrate. In the FDTD simulation software \cite{Ansys}, we have applied a periodic boundary condition (PBC) along the x and y dimensions (the x-y plane is chosen to be the MSR plane) and perfectly matched layer (PML) boundaries along z directions. In this work, numerical calculations are relevant to the far-field spectra; thus, one unit cell of the MSR under PBCs along the x-y plane was sufficient to simulate. Applied MSR parameters are nanohole etch depth 200 nm, etch diameter 120 nm and periodicity 300 nm; the same parameters were used for simulations in our previous work \cite{OurAdomPaper}. Scattering field responses were recorded by a planar field monitor installed along the MSR plane, inside the nanohole, where the local field enhancement is maximum. For the purpose of the presentation, we have used symmetric arrangements ($2\times2$ array) of nanoholes and recorded field profiles for individual scattering fields of induced multipole moments.\\ A plane wave source was used to excite the MSR from the top at normal incidence; the field monitor placed at the plane of the nanohole etch depth records local scattering responses. The dipole source was placed at the nanohole etch depth for maximal coupling, as we learned from our previous work \cite{OurAdomPaper}.\\
All the results, for example, intensities, are duly normalized with respect to the corresponding values obtained for the same simulation setup with corresponding sources but in free space. This general method helps eliminate any source dependence (e.g., source power, amplitude, position, etc.), and the scattering results are ready to compare since they only depend upon the pure response from the system \cite{OurAdomPaper}. 

\section{Discussion on Results}\label{sec: SI_MomentDecomposition}
\subsection{Generation of discrete dark modes}
Among the local scattering responses of induced moments, only the field generated by ED reaches the far field of the outcoupling direction. Both ED and MD-generated fields follow a distance relation $\sim\frac{1}{r^3}$, and the TD-generated field follows a distance relation $\sim\frac{1}{r^4}$ in their near fields. We can understand that following Jackson's textbook \cite{jackson_electrodynamics}. Fields generated by ED are,
\begin{equation}\label{eq: ED_fields}
\begin{aligned}
& \mathbf{H}=\frac{c k^2}{4 \pi}(\mathbf{n} \times \mathbf{p}) \frac{e^{i k r}}{r}\left(1-\frac{1}{i k r}\right) \\
& \mathbf{E}=\frac{1}{4 \pi \epsilon_0}\left\{k^2(\mathbf{n} \times \mathbf{p}) \times \mathbf{n} \frac{e^{i k r}}{r}+[3 \mathbf{n}(\mathbf{n} \cdot \mathbf{p})-\mathbf{p}]\left(\frac{1}{r^3}-\frac{i k}{r^2}\right) e^{i k r}\right\}
\end{aligned}
\end{equation}
\begin{figure}[!ht]
    \centering
    \includegraphics[width=0.85\linewidth]{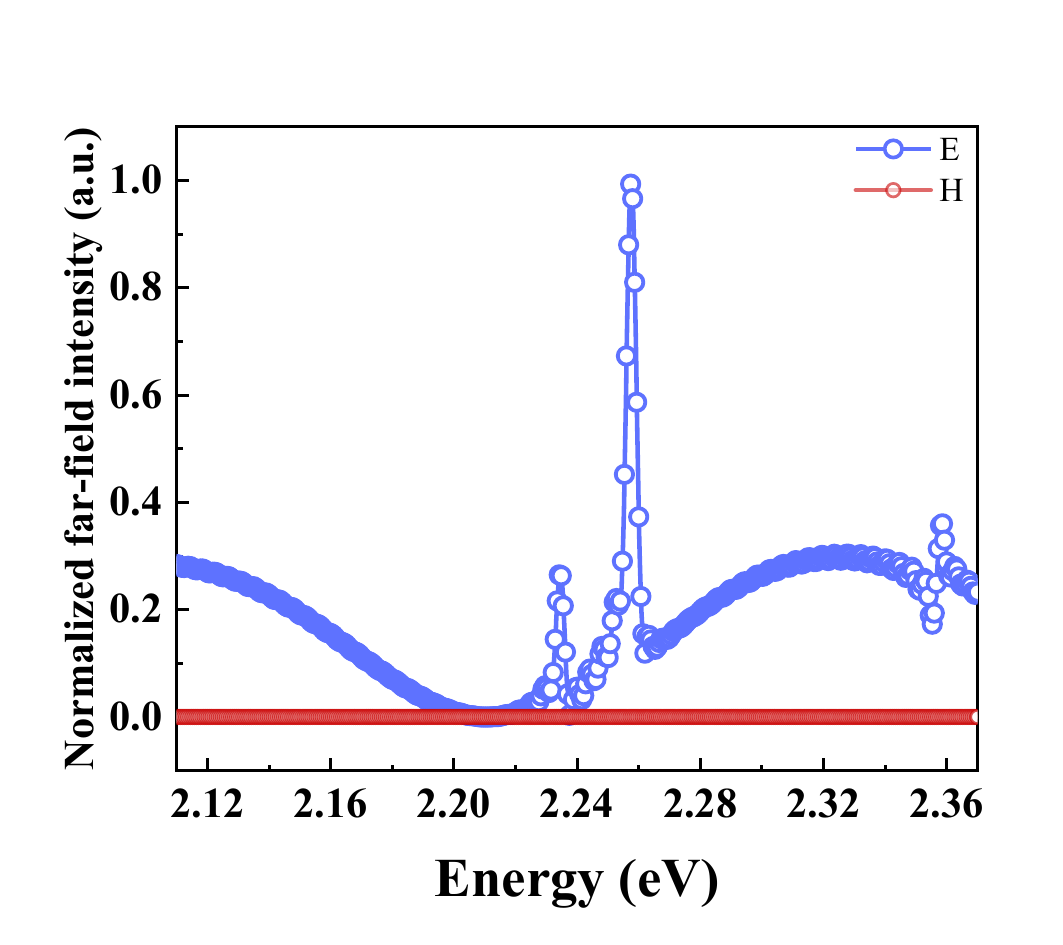}
    \caption{The plot the far-field spectra of the electric and the magnetic field scattering from the MSR. Plot is normalized with respect to the maximum of the electric field resonance. The corresponding value of the H-field spectrum lies around $7\times10^{-6}$ in arbitrary units.}
    \label{fig: ff_E_H}
\end{figure}

\begin{figure}[!ht]
    \centering
    \includegraphics[width=0.42\textwidth]{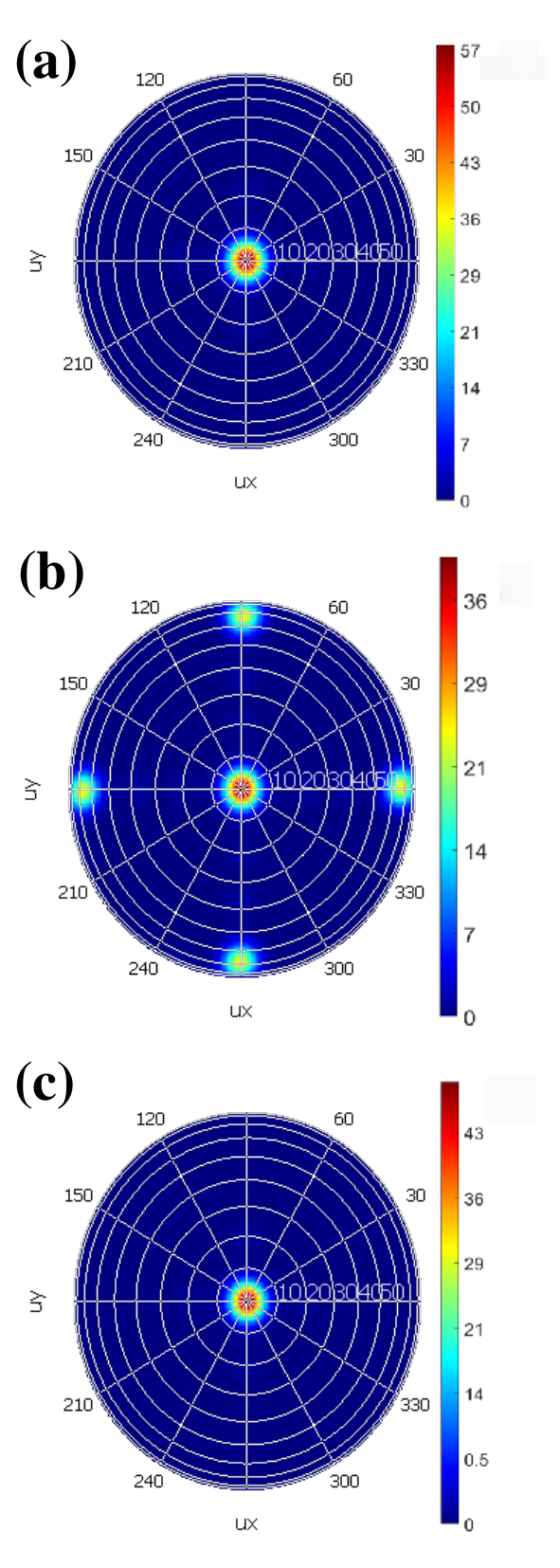}
    \caption{Far field projection of the MSR scattering. \textbf{(a)} and \textbf{(c)} show the far-field projection of radiation at frequencies just detuned from the resonance. \textbf{(b)} shows the radiation projection at resonance. Guided modes at the azimuthal angles for perpendicular edges are seen; the MSR has a square symmetric geometry.}
    \label{fig: SM_farfield}
\end{figure}

In the far field, these fields take the limiting form;
\begin{equation} \label{eq: ED_ff}
\begin{aligned}
& \mathbf{H}=\frac{c k^2}{4 \pi}(\mathbf{n} \times \mathbf{p}) \frac{e^{i k r}}{r} \\
& \mathbf{E}=Z_0 \mathbf{H} \times \mathbf{n}
\end{aligned}
\end{equation}
$Z_0=\sqrt{\mu_0 / \epsilon_0}$ is the impedance in free space. Since the induced ED moments $\textbf{p}$ are along the MSR plane,and the outcoupling direction $\textbf{n}$ is along the normal to the MSR, therefore the H field generated by ED is azimuthally oriented w.r.t dipoles. Total power radiated is given by,
\begin{equation}\label{eq: ED_power}
\frac{d P}{d \Omega}=\frac{c^2 Z_0}{32 \pi^2} k^4|\mathbf{p}|^2 \sin ^2 \theta
\end{equation}
Angle $\theta$ is measured from the direction of $\textbf{p}$.\\
Similarly, we obtain fields for an induced MD at the MSR plane;
\begin{equation}\label{eq: MD_fields}
\begin{aligned}
    & \mathbf{H}=\frac{1}{4 \pi}\left\{k^2(\mathbf{n} \times \mathbf{m}) \times \mathbf{n} \frac{e^{i k r}}{r}+[3 \mathbf{n}(\mathbf{n} \cdot \mathbf{m})-\mathbf{m}]\left(\frac{1}{r^3}-\frac{i k}{r^2}\right) e^{i k r}\right\}\\
    & \mathbf{E} = -\frac{Z_0}{4 \pi} k^2 (\mathbf{n} \times \mathbf{m}) \frac{e^{ikr}}{r} \left(1 - \frac{1}{ikr}\right)
\end{aligned}
\end{equation}
Where $\mathbf{m} = \int \mathbf{M} d^3x = \frac{1}{2} \int (\mathbf{x} \times \mathbf{J}) d^3x$ is the induced magnetic moment for the induced current $\textbf{J}$.
The total power radiated by an MD has an identical expression to Eq.\ref{eq: ED_power}, replacing $\textbf{p}$ with $\textbf{m}$. It is evident that the $\textbf{m}$ is perpendicular to the MSR plane here. Thus, the total power radiated by MD is along the azimuthal plane \cite{jackson_electrodynamics}.\\
The circulation of the current $\textbf{J}$ along the surface of a toroid having its axis aligned with the normal to the MSR plane generates the toroidal dipole moment \cite{TD_radiation1};
\begin{equation} \label{eq: TD_J}
    \textbf{J} = \vec{\nabla}\times\vec{\nabla}\times\textbf{T}
\end{equation}
$\textbf{T}$ is the induced toroidal moment aligned along the normal to the MSR. Fields generated by TD reaching the far field are unusual until the source is in a high-index dielectric material. Here, the MSR is made of SiN; therefore, the induced TD fields can reach the far field. The far-field relation is again the same as Eq.\ref{eq: ED_power} for the TD. Evidently, radiation power for TD follows the same rule as MD's, since their moments are parallel \cite{TD_radiation1, TD_radiation2}. Thus, the discrete guided modes of the MSR are nothing but the radiations of induced MDs and TDs. These are seen in the far field projection of radiation shown in Fig.\ref{fig: SM_farfield} for the simulated square symmetric MSR. Discrete guided modes are only present at resonance, and guided modes are at the azimuthal plane, emerging from the MSR edges, perpendicular to the corresponding sides. Figure \ref{fig: ff_E_H} shows the plot for the electric and magnetic field scattering from the MSR. The plot is normalized with respect to the maximum of the electric field intensity. The corresponding magnetic field intensities are lying around $7\times10^{-6}$, thus showing that such modes are indeed 'dark' along the far-field direction. However, from Figure \ref{fig:schematic} of the main text, it is evident that the MSR has $R+T<1$, but it will be equal to 1 when the guided modes are included, which are discrete symmetric modes.

\subsection{Discussion on the resonance linewidth}
\begin{table}[!ht]
\caption{\label{table_2eta} Data for the comparison of imaginary phase term 2Im$(\eta)$ and resonance linewidth $\Gamma$ obtained from Eq.(\ref{eq:7} - \ref{eq:11}) in the main text for all sources and interferences at $\omega_d$.}
\begin{ruledtabular}
\begin{center}
\begin{tabular}{cccc}
\textrm{Excitation}&
\textrm{Interference}&
\multicolumn{1}{c}{\textrm{2Im$(\eta)$}}&{\textrm{$\Gamma$}}\\
\textrm{source}&
\textrm{between}&
\multicolumn{1}{c}{\textrm{}}& {\textrm{(a.u.)}}\\
\colrule
Planewave & ED-MD & 59.6 & 29.94\\
 & ED-TD & 22.18 & 11.15\\
V dipole & ED-MD & 16.53 & 6.21\\
 & ED-TD & 1400.83 & 526.33\\
H dipole & ED-MD & $6.45\times10^4$ & 32330.63\\
 & ED-TD & $4.52\times10^3$ & 2247.46\\
\end{tabular}
\end{center}
\end{ruledtabular}
\end{table}
Intrinsic damping and corresponding resonance width due to the interference between the bright and dark mode [Eq.(\ref{eq:7}) and Eq.(\ref{eq:8}) in the main text] also find a change for the near-field interaction of the embedded dipole with the MSR compared to the planewave excitation of MSR. Table \ref{table_2eta} summarizes all these data. For both interferences, total $\Gamma$ and total 2Im$(\eta)$ have increments compared to their total values in planewave excitation. 2Im$(\eta)$ has the opposite effect to $\Gamma$. Data in Table \ref{table_2eta} show that 2Im$(\eta)$ increases more than $\Gamma$ when excitation was changed from planewave to embedded dipoles. Thus, the far-field spectra find a narrowing in their linewidth due to the change in the MSR excitation.

    \label{fig: SM_MSR_wk}
\twocolumngrid
\bibliography{apssamp}

\end{document}